\documentstyle[12pt,a4,psfig]{article}
 
\def\Pom{{I\!\!P}}

\begin{document}
\date{June 18, 1996}

\title{\vspace*{-0.5cm}
Diffraction dissociation, an important background to  
photon-photon collisions via heavy
ion beams at LHC }


\author{R.~Engel\\[2mm]
{\it Institut f\"ur Theoretische Physik,}\\
{\it Universit\"at Leipzig, D--04109 Leipzig, Germany}\\[1mm]
{\it and Universit\"at Siegen,}\\
{\it Fachbereich Physik, D--57068
Siegen, Germany }\\[5mm]
M.A.~Braun\thanks{Visiting professor IBERDROLA.
Permanent address: Department of High Energy Physics, University
of St.~Petersburg, 198904 Russia},
C.~Pajares and J.~Ranft\\[2mm]
{\it Departamento de F\'\i sica de Part\'\i culas,}\\
{\it Universidade de
Santiago de Compostela,}\\
{\it E--15706 Santiago de Compostela, Spain}
}

\maketitle

\begin{abstract}
The cross sections for the production of hadrons in quasi-real
photon-photon collisions in proton-proton and heavy ion interactions 
are compared with the corresponding cross sections
for central  diffraction and for photon-pomeron collisions. 
The signatures,
heavy ions or protons with only slightly changed momenta
together with two large rapidity gaps and a cluster of produced
hadrons in the central region, are nearly identical in  all
three 
processes. Therefore, it will be rather difficult to distinguish
the  reactions experimentally. It is found, that 
central diffraction is the dominant process in collisions of
protons, light and medium-heavy ions. The photon-pomeron and 
photon-photon
processes have quite similar cross sections 
in collisions of heavy ions like lead.
\end{abstract}

\vspace*{-20cm}
\begin{flushright}
Preprint US--FT / 18--96, SI 96-04
\end{flushright}
\vspace*{20cm}

\clearpage

\section{Introduction}
Double-photon exchange as well as double-pomeron exchange and
photon-pomeron
reactions are characterized by large rapidity gaps separating
the remnants of the colliding hadrons or heavy ions from the
particles produced in the central rapidity region. This feature
is likely to give a considerable reduction of the background to
the studied reaction.

Photon-photon collisions in proton-proton or heavy ion reactions have been
discussed repeatedly 
\cite{Baur88a,Grabiak89a,Papageorgiu90a,Vidovic93a,Hencken95a} 
as a very attractive reaction
channel. Processes studied include Higgs 
production 
\cite{Drees89a,Papageorgiu89b,Greiner91a,Greiner93a,Papageorgiu95a} and
production of SUSY particles \cite{Greiner93a,Ohnemus94}. 
Also the ALICE Collaboration discusses this option for the
heavy ion experiment at the Large Hadron
Collider (LHC) of CERN \cite{ALICETP}.

Central diffraction (CD) and especially hard central 
diffraction is recognized as an interesting  new reaction channel
at future hadron and heavy ion colliders at least since the first model for
diffractive hard scattering due to Ingelman and Schlein
\cite{Ingelman85} and the  pioneering
experiment of the UA8 Collaboration \cite{Bonino88,Brandt92}.
For example, Higgs production in this channel was discussed 
in Refs.~\cite{Schaefer90,Bialas91,Fletcher93,Ellis90b}
and the production of heavy
flavors was studied 
in Refs.~\cite{Streng86a,Bialas92,Bialas94a,Heyssler96}.
The resonance and glueball production in photon-photon
collisions and in central diffraction was compared in 
Refs.~\cite{Natale94a,Natale95a}.

Here we will discuss both processes  and in addition
photon-pomeron collisions for the same reactions 
in the framework of the two-component 
Dual Parton Model (DPM) \cite{Capella87b,Aurenche92a,Bopp94a} 
and in particular in
the version as implemented in the DPM event generator {\sc
Phojet} \cite{Engel95a,Engel95d}.
We restrict the paper to the calculation of cross sections for
hadron production in the three processes, in a later step we might
extend the calculation to  cross sections for heavy flavor or new
particle production.

Hadronic photon-photon collisions within this framework were
already studied and compared to data in a recent paper by Engel and
Ranft \cite{Engel95d} mainly for photon-photon collisions at
electron-positron colliders. Here we report about the extension
of the {\sc Phojet} model 
 to photon fluxes as expected at proton-proton and
heavy ion colliders.

Hard diffraction within the
two-component DPM was first studied 
and compared to UA8 data and data from HERA
by Engel, Ranft, and Roesler \cite{Engel95c}. 
In the present paper we will 
calculate central  diffraction cross sections  as well as
photon-pomeron cross sections for
proton-proton and heavy ion collisions. 
Using these cross sections we compare 
basic features of the
three reactions, double-pomeron, double-photon and
photon-pomeron interactions
at proton-proton and heavy ion colliders, which have very similar 
experimental signatures.  A second more
detailed study of particle production in double-pomeron and 
double-photon scattering,
also using the {\sc Phojet} event generator will
be published soon \cite{Engelpopo}.

In Section 2 we describe the models used. Section 3 gives a
detailed comparison of the cross sections and hadron
distributions expected in the three channels. A summary is given in
Section 4 and in the Appendix we collect all the details about
photon flux calculations in heavy ion reactions as used in this paper.

\section{The Models}
\subsection{The event generator {\sc Phojet}}

The realization of the DPM with a soft and a hard
component in {\sc Phojet} is similar to the event 
generator {\sc Dtujet}-93
\cite{Aurenche92a,Aurenche94a} simulating $pp$ and
p-$\rm \bar p$ collisions up to very high energies. {\sc Phojet}
is applicable to collisions of stable hadrons as well as of
photons. 
Here we give only a short summary on the physics of
{\sc Phojet}, for more detailed descriptions we refer to
\cite{Engel95a,Engel95d,Engelpopo,Engel94c,Engel95b}.

The interactions of hadrons and
the hadronic fluctuations of the photon are described
within the Dual Parton Model in terms of
reggeon and pomeron exchanges. 
The physical photon state is treated as a
superposition of a ``bare photon'' and virtual hadronic states having the
same quantum numbers as the photon.
For soft processes, photon-hadron duality is used.
The energy-dependences of the reggeon and
pomeron amplitudes are assumed to be the same for all hadronic processes.
Therefore, data on hadron-hadron and photon-hadron cross sections
can be used to determine the parameters necessary to describe soft 
photon-photon interactions.
Inelastic interactions are
subdivided into processes involving only {\it soft}
processes and all the other processes with at least one large momentum
transfer ({\it hard} processes) by applying
a transverse momentum cutoff $p_\perp^{\mbox{\scriptsize cutoff}}$ to
the partons.
On Born-graph level, for example, the photon-photon cross section is built
up by: {\bf (i)} soft reggeon and pomeron exchange,
{\bf (ii)} hard double-resolved photon-photon interaction,
{\bf (iii)} hard single-resolved interactions, and
{\bf (iv)} hard direct interactions.
The Parton Model 
calculations of the hard processes have been done using
the leading order GRV parton distribution functions for the 
proton \cite{GRV92a} and the photon \cite{GRV92b}.

The amplitudes corresponding to the one-pomeron exchange between the
hadronic fluctuations of the photon are
unitarized applying a two-channel eikonal formalism similar to
\cite{Aurenche92a}. 
In impact parameter representation, the eikonalized scattering amplitude
for resolved photon  interactions has the structure
\begin{equation}
a_{\mbox{\scriptsize res}}(s,B) = \frac{i}{2} 
\left(\frac{e^2}{f^2_{q\bar q}}\right)^2\;
\left( 1 - e^{-\chi(s,B)}\right)
\label{eff-amp}
\end{equation}
with the eikonal function
\begin{equation}
\chi (s,B)=\chi_{S}(s,B)+\chi_{H}(s,B)+\chi_{D}(s,B)+\chi_{C}(s,B).
\end{equation}
Here, $\chi_{i}(s,B)$ denote the contributions from the different
Born graphs: (S) soft part of the pomeron and reggeon, (H) hard part 
of the pomeron
(D) triple- and loop-pomeron graphs, (C) double-pomeron graphs.
In {\sc Phojet}  the last two terms are
included in the unitarization whereas in {\sc Dtujet} they are
taken in  lowest order.
The amplitude for hadron-hadron interactions has the same structure as 
(\ref{eff-amp}) but without the couplings
$e/f_{q\bar q}$ of the photon to the hadronic fluctuations.

The probabilities to find a photon in one of the generic hadronic 
states, the reggeon and pomeron coupling constants, and the 
effective reggeon and pomeron intercepts cannot be 
determined by basic
principles. These quantities are treated as free parameters and determined
by cross section fits \cite{Engel95a}. In Refs.~\cite{Engel95a,Engel95d}
the model predictions for the proton-proton, photon-proton, and
photon-photon cross sections are shown and compared to data.

The probabilities for the different final
state configurations are calculated from the discontinuity of the
scattering amplitude (optical theorem) which 
can be expressed as a sum of graphs with $k_c$ soft pomeron cuts, $l_c$
hard pomeron cuts, $m_c$ triple- or loop-pomeron cuts, and $n_c$
double-pomeron cuts by applying the Abramovski-Gribov-Kancheli cutting
rules \cite{Abramovski73,TerMartirosyan73}{}. 
In impact parameter space one gets for the inelastic cross
sections
\begin{eqnarray}
\sigma (k_{c},l_{c},m_{c},n_{c},s,B)&=&\nonumber\\
& &\hspace*{-3.2cm}\frac{(2\chi_{S})^{k_{c}}}{k_{c}!}
\frac{(2\chi_{H})^{l_{c}}}{l_{c}!}\frac{(2\chi_{D})^{m_{c}}}{m_{c}!}
\frac{(2\chi_{C})^{n_{c}}}{n_{c}!}\exp[-2\chi (s,B)].
\label{cutpro}
\end{eqnarray}
Since the triple-, loop-, and double-pomeron graphs are objects 
involving several pomerons, a further resummation is done
\cite{Aurenche92a,Engel95b} to allow the probability interpretation of
Eq.~(\ref{cutpro}).

For pomeron
cuts involving a hard scattering, the complete parton kinematics and
flavors/colors are sampled according to the Parton Model using a method
similar to \cite{Hahn90}{}, extended to direct processes. 
For pomeron cuts without hard large
momentum transfer, the partonic interpretation of the Dual Parton Model
is used: photons or mesons are split into a quark-antiquark pair whereas 
baryons are approximated by a quark-diquark pair.
The longitudinal momentum
fractions of the soft partons
are given by 
Regge asymptotics \cite{Capella80b,Kaidalov82a}.
The transverse momenta of the soft partons are sampled from an 
exponential distribution in order to get a smooth transition between
the transverse momentum distributions of the
soft constituents and the hard scattered partons.

In diffraction dissociation or double-pomeron scattering, 
the parton configurations are generated using the
ideas of the two-component DPM applied to pomeron-hadron,
photon-pomeron and pomeron-pomeron
scattering processes (see \cite{Engel95c} and references therein). 
For the parton densities in the pomeron we use the CKMT parametrization
\cite{Capella96a}.

Finally, the fragmentation of the sampled partonic final states is done
by forming color neutral strings between the partons according to the 
color flow. In the limit of large number of colors in QCD, 
this leads to the two-chain configuration
characterizing a cut pomeron and a one-chain system for a cut reggeon.
The chains are fragmented using the 
Lund fragmentation code {\sc Jetset} 7.3
\cite{Bengtsson87}.


\subsection{Diffraction dissociation in the two-component Dual Parton Model}

\subsubsection{Single diffraction dissociation}

\begin{figure}[thb] \centering
\hspace*{0.25cm}
\psfig{file=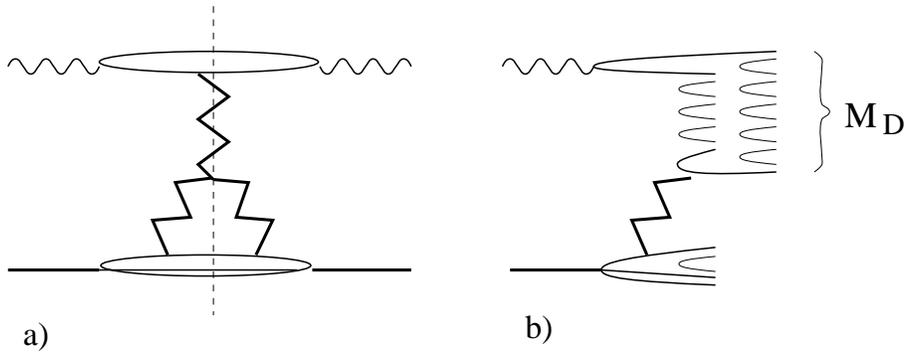,width=120mm}
 \vspace*{1cm}
\caption{
\label{hmdiff-g}
High-mass photon diffraction dissociation: diffractive cut of the
triple-pomeron graph a) and the corresponding chain system b)}
\end{figure}

In the model, low- and high-mass diffraction dissociation is distinguished.
Low-mass diffraction dissociation is described by assuming the scattering
of a superposition of resonances with the quantum numbers of the
dissociating particle \cite{Good60}. For simplicity, only one generic
resonance is considered in the model. In this case, the cross section
of diffractive processes with low-mass final states ($M^2_D < 2$ GeV$^2/c^4$)
can be calculated using a two-channel eikonal
formalism \cite{Kaidalov79,Aurenche92a,Engel95a}.
In the limit of large diffractive masses
$s \gg M^2_D \gg s_0$ and $M_D^2 \gg t$ with $s_0 \approx 1$ GeV$^2$,
the data can be understood in
terms of the triple-pomeron graph \cite{Kaidalov79,Roesler93}.
Here, $M_D$ denotes the diffractively produced mass and $t$ is the squared
four-momentum transfer.
In Fig.~\ref{hmdiff-g}, the triple-pomeron graph and the corresponding
diffractive cut for diffractive dissociation
of particle $A$  and quasi-elastic deflection of particle $B$
is shown.
Assuming multiperipheral kinematics of the pomeron cut final states,
the rapidity gap is approximately given by
$\eta_{\rm gap} \approx \ln(s/M_D^2)$.
It is convenient to characterize the final state using the Feynman $x_F$ of
the elastically scattered particle $B$
\begin{equation}
x_B = -\left(1 - \frac{M_D^2}{s}\right)
\end{equation}
which leads to
\begin{equation}
M_D^2 \approx (1-|x_B|) s
\hspace*{0.5cm} \mbox{and} \hspace*{0.5cm}
\eta_{\rm gap} \approx \ln\left(\frac{1}{1-|x_B|}\right)
\end{equation}
In Born-graph approximation neglecting rescattering effects,
the double-differential
cross section for high-mass diffraction dissociation reads
\begin{eqnarray}
\frac{d^2\sigma_{\rm TP}}{dt\,dM_D^2} &=& \frac{1}{16 \pi}
\left(g^0_{A\Pom}\right)^2\ g^0_{\Pom\!\Pom\!\Pom}\ g^0_{B\Pom}
\left(\frac{s}{s_0}\right)^{2\Delta_\Pom}
\left(\frac{s_0}{M_D^2}\right)^{\alpha_\Pom(0)}
\nonumber\\
& & \times
\exp\left\{ \left(b_{A\Pom} + b_{\Pom\!\Pom\!\Pom} + 2 \alpha^\prime_\Pom(0)
\ln\left(\frac{s}{M_D^2}\right)\right)\ t\right\}.
\label{triple-Pom}
\end{eqnarray}
with $\Delta_\Pom = \alpha_\Pom(0)-1$, $\alpha_\Pom(0)$ beeing the 
pomeron intercept.
The coupling constants are parametrized by
\begin{eqnarray}
g_{i\Pom}(t) &=& g^0_{i\Pom} \exp\left(\frac{1}{2}b_{i\Pom} t\right)
\hspace*{0.5cm} i=A,B\\
g_{\Pom\!\Pom\!\Pom}(t_1,t_2,t_3)
&=& g^0_{\Pom\!\Pom\!\Pom} \exp\left(\frac{1}{2}
b_{\Pom\!\Pom\!\Pom}(t_1+t_2+t_3)\right).
\nonumber\\
\end{eqnarray}

In the following we will restrict the diffractively produced mass 
according to an experimentally motivated cut on the Feynman $x_F$ 
of the elastically scattered particle \cite{Streng86a}
\begin{equation}
|x_B| > c, \hspace*{1cm} M_{D,\rm min}^2 \le M_D^2 \le (1-c) s,
\end{equation}
with $c = 0.9\dots0.97$.
The integration over $t$ and $M_D^2$ can be performed analytically 
\cite{Capella74}
\begin{eqnarray}
\sigma_{\rm TP} &=& \frac{1}{16 \pi} \left(\frac{s}{s_0}\right)^{\Delta_\Pom}
\frac{\left(g^0_{A\Pom}\right)^2\ g^0_{\Pom\!\Pom\!\Pom}\ g^0_{B\Pom}}{
2 \alpha^\prime_\Pom(0)}
\nonumber\\
&\times&\exp\left\{ - \Delta_\Pom 
\frac{b_{A\Pom}+b_{\Pom\!\Pom\!\Pom}}{2 \alpha^\prime_\Pom(0)} \right\}
\nonumber\\
&\times& \Bigg[ Ei\left( \Delta_\Pom 
\left( \frac{b_{A\Pom}+b_{\Pom\!\Pom\!\Pom}}{2 \alpha^\prime_\Pom(0)}
+\ln\frac{s}{M^2_{D,\rm min}}\right) \right)
\nonumber\\
& &\hspace*{3mm}
- Ei\left( \Delta_\Pom 
\left( \frac{b_{A\Pom}+b_{\Pom\!\Pom\!\Pom}}{2 \alpha^\prime_\Pom(0)}
+\ln\frac{1}{1-c}\right) \right) \Bigg]\ ,
\end{eqnarray}
where $Ei$ denotes the second exponential integral function.

Elastic and inelastic rescattering
effects decrease the cross section as given by Eq.~(\ref{triple-Pom})
considerably \cite{Capella76,Gotsman93}.
This suppression is estimated using the eikonal model as described in the
previous section.
In impact parameter representation, the experimentally observable
cross section of diffraction dissociation 
follows from (see Eq.~(\ref{cutpro}))
\begin{equation}
\sigma^{d}(s,B) = \sigma(k_c=0,l_c=0,m_c=1,n_c=0;s,B)\ .
\end{equation}

The free parameters of the model (coupling constants, pomeron intercept,
slope parameters) are determined by a global fit to data on
total, elastic and diffractive cross sections as well as data on elastic slope
of $pp$, $p\bar p$ and $\gamma p$ interactions \cite{Engel95a}.
Assuming that soft hadronic interactions in hadron-hadron,
photon-hadron, and photon-photon interactions can be described by
the exchange of one universal object, the pomeron, all these data can
be combined to increase the predictive power of the model.

Fig.~\ref{hmdiff-g} suggests the interpretation of photon diffraction
in terms of photon-pomeron scattering.
However, since the pomeron is only a theoretically
introduced object to describe the some features of hadronic
high-energy scattering, it is not possible to consider photon-pomeron
scattering without the corresponding hadron where the pomeron couples to.

Finally, it should be mentioned that
the model results can be compared with the new measurements
on the cross section of photon single diffraction dissociation at HERA.
This is important since the cross section  on photon diffraction
dissociation enters directly the predictions on the photon-pomeron
cross sections. For $\sqrt{s}_{\gamma p}= 200$ GeV, the model predicts
a cross section of 19$\mu$b to be compared with $23\pm11\mu$b \cite{Aid95b}.
Furthermore, in Ref.~\cite{Aid95b},
some model results are shown together with data
on photon diffraction obtaind by the H1 Collaboration, finding reasonable 
agreement.

\subsubsection{Central diffraction}

A direct consequence of the interpretation of diffraction as
pomeron-particle scattering is the prediction of the existence of
pomeron-pomeron scattering (double-pomeron
scattering).
Double-pomeron cross sections in hadron-hadron collisions were
first calculated on the basis of single diffractive measurements
and Regge theory by Chew and Chew \cite{Chew74} and by Kaidalov
and Ter-Martirosyan \cite{Kaidalov74a}. Using different models, these 
calculations have been continuously 
extended to partial cross section estimations (see 
for example Refs.~\cite{Berera95a,Pumplin93a} and references therein).

The calculation of double-pomeron cross sections is subject to
considerable uncertainty. There are several reasons for
this: 

\noindent
(i) The cross section is proportional to the square of the
triple-pomeron coupling constant $g_{\Pom\!\Pom\!\Pom}$, 
which is not very well known.

\noindent
(ii) The double-pomeron cross section differs considerably, if this
term is included into an unitarization procedure for all
hadronic cross sections or not. For example,
it was discussed recently \cite{Gotsman95a}, that the effects of
shadowing decrease the double-pomeron cross sections at LHC
energies by a factor of 5. The cross section estimates as given for example in
\cite{Kaidalov74a,Streng86a} do not include any unitarization whereas
in our approach the double-pomeron cross sections are included in 
the unitarization.

\noindent
(iii) The energy dependence of the double-pomeron cross section
might differ considerably when using either a critical or
a supercritical pomeron intercept $\alpha_\Pom(0)$. 
In \cite{Kaidalov74a,Streng86a} a
critical pomeron intercept $\alpha_\Pom(0) = 1$ was used. 
Here we use, consistent with high-energy cross section measurements,
a supercritical intercept $\alpha_\Pom(0) > 1$.

\noindent
(iv) On a more practical level these cross sections depend on
the cuts applied to the centrally produced cluster of
particles and on the rapidity gaps demanded by the experimental
triggers.

\noindent
(v) In CD, one or both of the incoming hadrons 
can be excited to a resonance $N^\star$.
In the following, for the comparison with the two-photon cross sections,
 we exclude from the calculation all cross section contributions involving
resonances.  Including these contributions would about double the cross
sections given below. 
These contributions would also change the rapidity gap demanded 
in experimental trigger conditions.

\begin{figure}[thb] \centering
\hspace*{0.25cm}
  \psfig{file=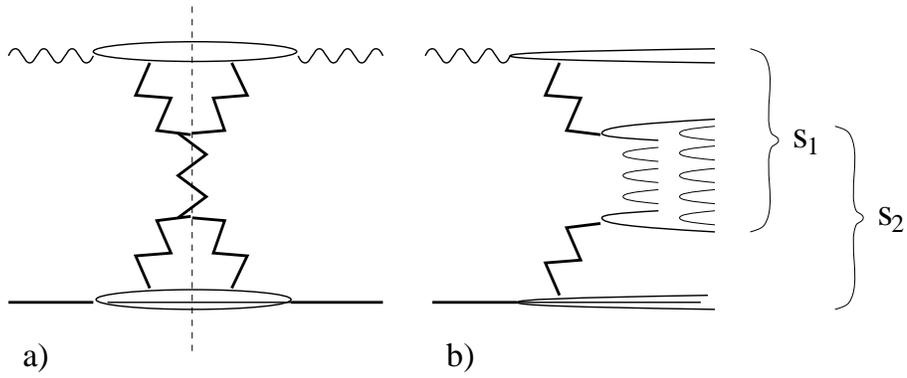,width=120mm}
 \vspace*{1cm}
\caption{
Central diffraction: diffractive cut of the double-pomeron graph a) 
and the corresponding chain system b).
\label{cdiff} 
}
\end{figure}

In the approximation of multiperipheral kinematics,
the mass $M_{cd}$ of the centrally produced diffractive system is
$M_{cd}^2 = (s_1 s_2)/s$ (see Fig.~\ref{cdiff}).
The rapidity gaps between the central system and the elastically
scattered particles can be approximated by
\begin{equation}
\eta_{{\rm gap} 1} \approx \ln\left( \frac{s}{s_2} \right)\hspace*{2cm}
\eta_{{\rm gap} 2} \approx \ln\left( \frac{s}{s_1} \right)\ .
\end{equation}
Denoting the Feynman $x_F$ of the elastically scattered particles $A$
and $B$ by $x_A$ and $x_B$ one gets \cite{Streng86a}
\begin{equation}
x_A = 1-\frac{s_2}{s}\hspace*{2cm}x_B = -\left(1-\frac{s_1}{s}\right)
\end{equation}
and
\begin{eqnarray}
& &M_{cd}^2 \approx (1-|x_A|)(1-|x_B|)s\nonumber\\
& &\eta_{{\rm gap} 1} \approx \ln\left( \frac{1}{1-|x_A|} \right)
\hspace*{1cm}
\eta_{{\rm gap} 2} \approx \ln\left( \frac{1}{1-|x_B|} \right)\ .
\end{eqnarray}

Within the framework of Gribov's Reggeon calculus, 
the amplitude of the double-pomeron
graph can be calculated (for more details see \cite{Kaidalov74a,Engelpopo}). 
From this one gets the cross section as function of $t_A$, $t_B$, $s_1$, and
$s_2$ ($t_A$ and $t_B$ denote the squared momentum transfer 
of particle $A$ and $B$). After integration over $t_A$ and $t_B$ 
the differential cross
section reads
\begin{eqnarray}
\frac{d\sigma_{\rm DP}}{ds_1 dM_{cd}^2} &=& \frac{1}{256 \pi^2}
\sigma_{\Pom\Pom}(M_{cd}^2) \left(\frac{s}{M_{
cd}^2}\right)^{2\Delta_\Pom} \frac{1}{M_{cd}^2} \frac{1}{s_1}
\nonumber\\
&\times&
\frac{\left( g^0_{A\Pom}\right)^2}{
b_{A\Pom}+b_{\Pom\!\Pom\!\Pom}+2\alpha_\Pom^\prime(0)\ln\left(\frac{s_1}{
M_{cd}^2}\right)} \nonumber\\
&\times&
\frac{\left(g^0_{B\Pom}\right)^2}{
b_{B\Pom}+b_{\Pom\!\Pom\!\Pom}+2\alpha_\Pom^\prime(0)\ln\left(\frac{s}{
s_1}\right)}
\end{eqnarray}
with
\begin{equation}
\sigma_{\Pom\!\Pom}(M_{cd}^2) = \left( g^0_{\Pom\!\Pom\!\Pom}\right)^2
\left(\frac{M_{cd}^2}{s_0}\right)^{\Delta_\Pom}\ .
\end{equation}

Applying the previously discussed cut on the Feynman 
$x_F$ of the elastically scattered hadrons
\begin{eqnarray}
& &x_A \ge c,\hspace*{2cm} |x_B| \ge c,\nonumber\\
& &M_{cd, \rm min}^2 \le M_{cd}^2 \le (1-c)^2 s
\label{xf-cuts}
\end{eqnarray}
the integration over $s_1$ can be performed
\begin{equation}
\frac{M_{cd}^2}{1-c} \le s_1 \le (1-c) s
\end{equation}
\begin{eqnarray}
M_{cd}^2\frac{d\sigma_{\rm DP}}{dM_{cd}^2} &=& \frac{1}{256\pi^2}
\sigma_{\Pom\Pom}(M_{cd}^2) \left( g^0_{A\Pom}\;
g^0_{B\Pom}\right)^2 
\nonumber\\
& &\hspace*{-1.3cm}
\times
\frac{1}{\alpha_\Pom^\prime(0)}
\left(\frac{s}{M_{cd}^2}\right)^{2\Delta_\Pom}
\nonumber\\
& &\hspace*{-1.3cm}
\times
\ln\left(
\frac{
b_{A\Pom}+b_{\Pom\Pom\Pom}+2\alpha_\Pom^\prime(0)\ln((1-c)s/M_{cd}^2)}{
b_{B\Pom}+b_{\Pom\Pom\Pom}+2\alpha_\Pom^\prime(0)\ln(1/(1-c))}\right)
\nonumber\\
& &\hspace*{-1.3cm}
\times \left(b_{A\Pom}+b_{B\Pom}+2b_{\Pom\Pom\Pom}
+2\alpha_\Pom^\prime(0)\ln(s/M_{cd}^2)\right)^{-1}\ .
\nonumber\\
\label{DP-final}
\end{eqnarray}
In order to calculate the cross section for CD,
the Born graph cross section (\ref{DP-final}) for double-pomeron scattering
is included in the eikonalization.  In impact
parameter representation, the CD cross section reads
(see Eq.~(\ref{cutpro}))
\begin{equation}
\sigma^{cd}(s,B) = \sigma(k_c=0,l_c=0,m_c=0,n_c=1;s,B)\ .
\end{equation}

\begin{figure}[thb] \centering
\hspace*{0.25cm}
\input{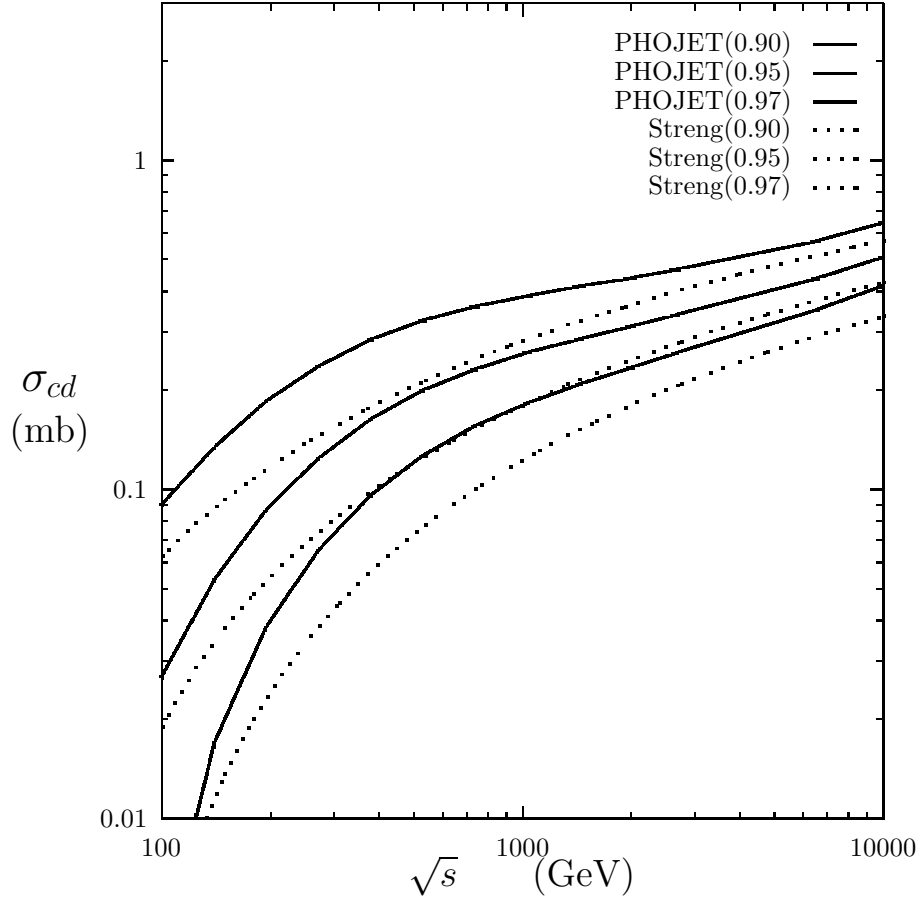}
 \vspace*{1cm}
\caption{The energy dependence of the central diffraction
cross section. We compare the cross section as obtained from
{\sc Phojet} with unitarization using a supercritical
pomeron with the cross section obtained by Streng
\protect\cite{Streng86a} without unitarization and with a critical
pomeron. Both cross sections are for the same three kinematical cuts:
$M_{cd} > $2GeV/c${}^2$ and $c =$0.90, 0.95 and 0.97.
Please note, to identify the cross sections, the cross sections
decrease with rising $c$.
\label{sigdpoecm}
}
\end{figure}

In Fig.~\ref{sigdpoecm} we compare as function of the energy 
the CD cross
sections in proton-proton collisions, 
which we obtain from {\sc Phojet} with the cross
section obtained by Streng \cite{Streng86a}. For both calculations
the same three kinematical cuts are used:
$M_{cd} > $2GeV/c${}^2$ and $c= $0.90, 0.95 and 0.97.
In {\sc Phojet} we use a supercritical pomeron with
$\Delta_{\Pom}$ = 0.08 whereas Streng \cite{Streng86a}
uses a critical Pomeron with $\Delta_{\Pom}$ = 0.
Note that the double-pomeron cross section grows in Born approximation
with $s$ like 
$\sim s^{2\Delta_\Pom}$. This rapid increase is damped 
in {\sc Phojet} by the unitarization procedure. At high energies, 
contributions from multiple interactions become important. 
The demanded rapidity gaps are filled with hadrons due to 
inelastic rescattering and the cross section for CD
gets strongly reduced. In contrast, Streng
calculates only the Born term cross section.
Figure~\ref{sigdpoecm}
illustrates the  differences obtained using different methods.
We stress, both methods use the measured
single diffractive cross sections to extract the triple-pomeron
coupling.

Finally, it should be mentioned that in a very recent work \cite{Armesto96a}
pomeron-pomeron cross sections have been studied in the perturbative 
BFKL-Bartels approach (see \cite{Armesto96a} and references therein). 
The predictions on cross sections found in this work are of the same order 
as the ones obtained here.


\subsection{Diffractive cross sections in collisions involving
nuclei}

\subsubsection{Central diffraction cross sections in heavy ion
collisions}

There are certain difficulties
in resolving the  CD cross section as a function
of its central mass $M_{cd}$ in hadron-nucleus and nucleus-nucleus 
collisions. In fact, consider the simplest 
case of $pd$ scattering
with only two possible inelastic $pN$ collisions. If in the first collision a
central mass $M_{1}$ is produced, in the second collision another central mass
$M_{2}$ is produced and their rapidities overlap, then it would not be possible
to distinguish such an event from a single $pN$ collision with the mass
$M_{1}+M_{2}$ produced in the center. Therefore hadron-nucleus and
nucleus-nucleus collisions, rigorously speaking, only give information on the
events with two fixed rapidity gaps, without specifying the exact nature of the
object produced in between.

However the situation improves if one takes into account that the CD events
have a very small probability, so that the described event with two central
masses produced is highly improbable. Then the typical CD event will be that
the central mass is produced in only one of the inelastic collisions, all
others belonging to elastic or low mass diffractive  events. In such a
case the $M_{cd}$-dependence of the CD cross section 
will evidently repeat that for $pp$ collisions.

An appropriate tool to calculate the CD cross section in hadron-nucleus and
nucleus collisions is the so-called 
"criterion $C$" \cite{Blankenbecler81,Pajares81b,Pajares85a,Braun91}. 
It has been known
since long ago that in the Glauber model the inelastic cross section is
screened only by itself. As it turns out, there are many other types of events
which are screened by themselves. These events have to possess a certain
property $C$ which satisfies the following requirement: Any superposition of
$NN$ events, in which at least one has property $C$, leads to the $hA$ or $AB$
event with property $C$. As a consequence, the only possibility to obtain an
$hA$ or $AB$ event without property $C$ should be that all $NN$ events do not
possess this property.

To translate the criterion $C$ into formulas, let
$\sigma^{in}_{AB}(\sigma)$ be the inelastic nucleus $A$-nucleus $B$
cross section considered as a function of the total nucleon-nucleon
cross section $\sigma$. Then the criterion $C$ tells that for the events
with the property $C$
\begin{equation}
\sigma^{c}_{AB}=\sigma_{AB}^{in}(\sigma^{c})
\end{equation} which  exactly means that they are shadowed only by themselves.

Passing to CD events, we have to apply the criterion $C$ twice. First choose
as events satisfying the criterion $C$ those with particles produced at least
in  one of the two fixed rapidity gaps which determine the central region.
Evidently any superposition of such  $NN$ events leads to an $AB$ event of
the same type. The complementary events are those in which no particle is
produced in any of the two gaps. For $NN$ collisions they include  elastic
events plus events in which particles are also produced in between the gaps and
in the low mass diffractive regions of the projectile and target, above the
upper gap and below the lower gap. Therefore the nucleon-nucleon cross section
$\sigma^{c}$ will be given by the difference
 \begin{equation}
\sigma^{c}=\sigma-\sigma^{el}-\sigma^{lmd}-\sigma^{cd}
\label{M2}
 \end{equation}
where $\sigma^{lmd}$ refers to the mentioned low-mass diffractive contribution.
$\sigma^{cd}$  denotes the central diffraction contribution (to which
also particles scattered elastically or low-mass diffraction contribute).
Note that the magnitude of the low-mass diffractive part depends on the 
choice of the experimental setup which defines the events satisfying the
criterion $C$. In particular, if these events are chosen to include all
low-mass diffractive contribution to the $NN$ cross section, then the term
$\sigma^{lmd}$ will not appear in (\ref{M2}) and the cross section $\sigma^{cd}$ will
include only events with some particle produced in between the gap plus
the projectile and target nucleons scattered elastically. In the following we
shall be interested precisely in such  CD events, having in mind that in the
experimental setup for the photon-photon interaction small values of the
momentum transferred to the colliding nuclei dominate, for which excitation of
nucleon resonances is prohibited. For that reason we assume that all events with
excitation of nucleon resonances are included in the events with the property
$C$. Then (\ref{M2}) simplifies to
\begin{equation}
\sigma^{c}=\sigma^{in}-\sigma^{cd}
\label{M3}
 \end{equation}

 Subtracting from the total  $AB$ cross section the one
with the described property $C$ we find
a cross section which is a sum  of the elastic cross section,
the cross section for the diffractive dissociation of the colliding nuclei and 
 the CD cross section in which the nuclei either stay intact or are
diffractively dissociated \begin{equation}
\sigma_{AB}^{el}+\sigma_{AB}^{difd}+\sigma_{AB}^{cd}=\sigma_{AB}^{tot}-
\sigma_{AB}^{in}(\sigma^{in}-\sigma^{cd})
\label{M4}
\end{equation}

Now we repeat this argument taking for events satisfying $C$ those in which any
particle is produced in the whole rapidity range spanned by the two gaps and
the rapidity interval in between. Such events also satisfy the conditions
implied by the criterion $C$. The complementary events are now those in which no
particle is produced in the described rapidity range at all. For $NN$ collisions
these are pure elastic events. For $AB$ they also include  nuclei diffraction
dissociation events. Similar to (\ref{M4}) we then get the well known formula
\begin{equation}
\sigma_{AB}^{el}+\sigma_{AB}^{difd}=\sigma_{AB}^{tot}-
\sigma_{AB}^{in}(\sigma^{in})
\label{M5}
\end{equation}

Subtracting (\ref{M5}) from (\ref{M4}) we obtain the desired CD cross section 
for $AB$ collisions
\begin{equation}
\sigma_{AB}^{cd}=\sigma_{AB}^{in}(\sigma^{in})-
\sigma_{AB}^{in}(\sigma^{in}-\sigma^{cd})
\label{M6}
\end{equation}

This general formula can be written in an explicit form for $hA$ collisions
where the explicit dependence $\sigma_{hA}^{in}(\sigma)$ is known. In the
Glauber model, for fixed impact parameter $B$,
\begin{equation}
\sigma_{hA}^{in}(\sigma)=1-(1-\sigma T(B))^{A}
\simeq 1-\exp ( -A\sigma T(B))
\label{M7}
\end{equation}
where $T(B)$ is the nuclear profile function. We then find  the CD
cross section
\begin{eqnarray}
\sigma_{hA}^{cd}&=&(1-(\sigma^{in}-\sigma^{cd})T(B))^{A}-
(1-\sigma^{in}T(B))^{A}\nonumber\\
&\simeq& \exp ( -A\sigma^{in}T(B))(\exp ( A\sigma^{cd}{T(B)}-1)
\label{M8}
\end{eqnarray}

From the derivation it is clear that this cross section refers to the total
probability to produce some particles between the two gaps and does not specify
the particles energy $M$. However, as mentioned, we can make use of the fact
that $\sigma^{cd}$ is very small. Then one approximately finds from (\ref{M8})
\begin{eqnarray}
\sigma_{hA}^{cd}&\simeq&\sigma^{cd}AT(B)(1-\sigma^{in}T(B))^{A-1}
\nonumber\\
&\simeq& \sigma^{cd}AT(B)\exp ( -A\sigma^{in}T(B))
\label{M9}
\end{eqnarray}
This cross section is linear in $\sigma^{cd}$. Therefore we can easily find
thep distribution in $M$
\begin{eqnarray}
\frac{d\sigma_{hA}^{cd}(M)}{dM}&=&\frac{d\sigma^{cd}(M)}{dM}
AT(B)(1-\sigma^{in}T(B))^{A-1}
\nonumber\\
&\simeq& \frac{d\sigma^{cd}(M)}{dM}AT(B)\exp ( -A\sigma^{in}T(B))
\label{M10}
\end{eqnarray}
As we observe, the absorption factor is universal and does not depend on $M$.

The application of this formalism to nucleus-nucleus collisions is hampered by
the absence of an explicit (and tractable) expression for
$\sigma^{in}_{AB}(\sigma)$. We shall use the well-known optical
approximation in which $\sigma^{in}_{AB}(\sigma)$ is given by the same
formula (\ref{M7}) with $A\rightarrow AB$ and an effective profile function for the
two colliding nuclei
\begin{equation} T_{AB}=\int d^2 B_{1}T_{A}(B_{1})T_{B}(B-B_{1})
\label{M11}
\end{equation}
where $B_{1}$ is the usual two dimensional impact parameter.
Our final formula for the CD cross section in $AA$ collisions then follows
from (\ref{M10})
\begin{equation}
\frac{d\sigma_{AA}^{cd}(M)}{dM}=\frac{d\sigma^{cd}(M)}{dM}A_{\rm eff}^{2}
\label{M12}
\end{equation}
where $A^2_{\rm eff}$ is the "effective" atomic number of the colliding 
nuclei defined by 
\begin{equation}
A^{2}_{\rm eff}=A^{2}\int d^{2}B
T_{AA}(B)\exp ( -A^{2}\sigma^{in}T_{AA}(B))
\label{M22}
\end{equation}

We compute $A^2_{\rm eff}$ using Woods-Saxon nuclear densities
\begin{equation}
\rho (r) = \frac{\rho _0}{1+e^{(r-r_0)/a}}
\end{equation}
with the standard parameter values \cite{Segre77} $r_0 =
1.14\; A^{1/3} $ fm and $a = 0.545$ fm and
and $\sigma^{in}=73$  mb, as
predicted in \cite{Engel95a} for $\sqrt{s}=6$ TeV. The values of
$A^2_{\rm eff}$ are given in Table~\ref{table1} .

One observes that
they are much smaller than $A^{2}$. 
Actually $A^{2}_{\rm eff}\sim A^{1/3}$, so that
even for very heavy colliding nuclei the CD cross section is only an order of
magnitude greater than for proton-proton collisions.

\subsubsection{Single  diffraction cross sections in hadron-nucleus
and photon-nucleus 
collisions}

The derivation given above can be applied also to single
diffraction with only one rapidity gap from the projectile or
target side. One only has to appropriately change the events
satisfying the criterion $C$. Then, instead of the central
diffractive nucleon-nucleon cross section $\sigma^{cd}$ in Eqs.~(\ref{M2}) 
-- (\ref{M10}) and  (\ref{M12}), the corresponding single
diffractive cross sections $\sigma^d$ will appear. In
particular, the final result for the nucleus-nucleus single
diffractive cross section will be
\begin{equation}
\sigma_{AB}^{d}=\sigma_{AB}^{in}(\sigma^{in})-
\sigma_{AB}^{in}(\sigma^{in}-\sigma^{d})
\label{M13}
\end{equation}
For $hA$ collisions we obtain instead of (\ref{M8}) and (\ref{M9})
\begin{eqnarray}
\sigma_{hA}^{d}&=& \exp ( -A\sigma^{in}T(B))(\exp ( A\sigma^{d}{T(B)}-1)
\nonumber\\
&\simeq& \sigma^{d}AT(B)\exp ( -A\sigma^{in}T(B))
\label{M14}
\end{eqnarray}
and an analogous formula for $d\sigma^d_{hA}/dM^2$ similar to
(\ref{M10}).

All of these formulae can be applied also for $\gamma\Pom$
interactions in heavy ion collisions.

The final formula for $\gamma A$ diffractive scattering is
similar to (\ref{M12})
\begin{equation}
\frac{d\sigma_{\gamma A}^{d}(M)}{dM}=\frac{d\sigma^{d}_{\gamma h}(M)
}{dM}A_{\rm eff}
\label{M15}
\end{equation}
with
\begin{equation}
A_{\rm eff}=A\int d^{2}B
T_{A}(B)\exp ( -A\sigma^{in}T_{A}(B))
\end{equation}
We calculate $A_{\rm eff}$ with the same input as for 
 $A^2_{\rm eff}$ in Eq.~(\ref{M22}) and present the values also
 in Table~\ref{table1}.
Actually also $A_{\rm eff}\sim A^{1/3}$, this is a behaviour
found before by Ranft and Roesler \cite{Ranft94b} and by Faessler
\cite{Faessler93a}.

\section{Comparing  hadronic
photon-photon interactions with  diffractive interactions 
in heavy ion collisions}

\subsection{Cross sections }

We calculate the cross sections $d\sigma /dM_{X}$ for the production
of a central cluster of hadrons with invariant mass $M_{X}$. This is
done for proton-proton collisions as well as for 
symmetrical heavy ion collisions with the 
same projectile and target ion. The calculations are done for
light (O), medium (Ca, Fe,  Ag) and heavy ions (Pb) with the energies
of the future
LHC hadron and heavy ion collider under construction at CERN.
The two ions most likely to be used in the LHC experiments
\cite{ALICETP} are Pb for obtaining the highest energy densities
and Ca, where a larger luminosity than with Pb can be obtained.
It was shown \cite{Brandt94a} that the luminosity in Ca-Ca
collisions could be up to a factor $10^4$ larger than in Pb-Pb
collisions. This gain more then compensates the loss in the two-photon 
flux in the Ca-Ca reaction.
($L_{\rm Pb-Pb} \approx 5 \times 10^{26} $ cm$^{-2}$s$^{-1}$, 
$L_{\rm Ca-Ca} \approx 5 \times 10^{30} $ cm$^{-2}$s$^{-1}$). 

For photon-photon collisions we use two different approaches for
calculating photon fluxes described in the Appendix.
These two
approximations give slightly different cross sections especially at
large masses $M_{X}$. 

\begin{figure}[thb] \centering
\hspace*{0.25cm}
\input{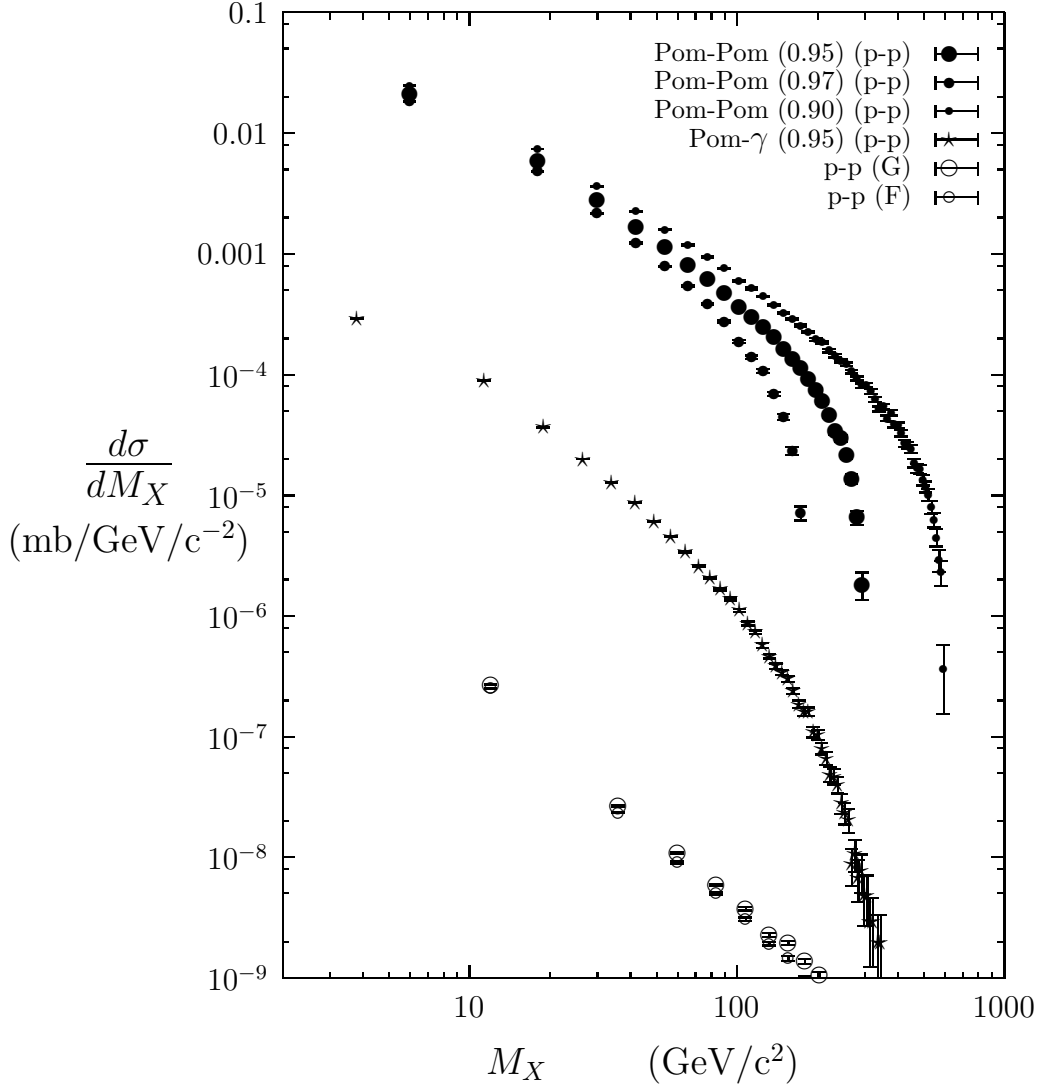}
 \vspace*{1cm}
\caption{
We compare the cross section for the
production of a hadronic cluster of invariant mass $M_{X}$ via
photon-photon interaction in
proton-proton collisions using two different 
methods (F) and
(G) to calculate the  photon flux, described in the
Appendix with the
corresponding cross section for the diffractive reactions. The
central diffraction cross sections (pomeron-pomeron collisions)
are given for three different
kinematical cuts 
\protect$M_{cd} >$ 2  GeV/c$^2$, 
$c =$ 0.90, 0.95 and 0.97.
The single diffraction photon-pomeron cross section is given
for $M_{\gamma\Pom} > $ 2 GeV/c$^2$ and c = 0.95.
\protect\label{hipssmblo}
}
\end{figure}

\begin{figure}[thb] \centering
\hspace*{0.25cm}
\input{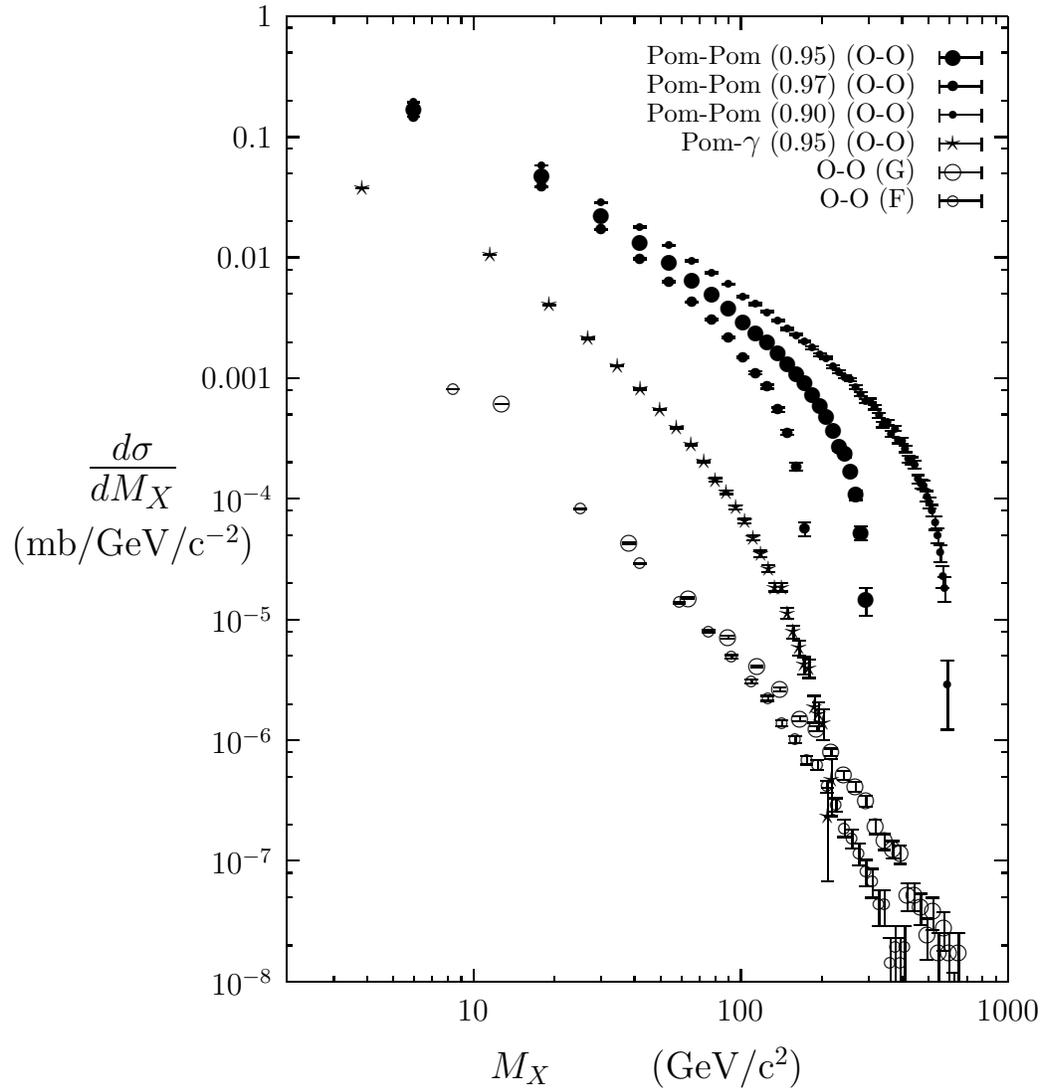}
 \vspace*{1cm}
\caption{
As Fig.~\protect\ref{hipssmblo} but for heavy ion reactions
O-O.
\label{hiossmblo}
}
\end{figure}

\begin{figure}[thb] \centering
\hspace*{0.25cm}
\input{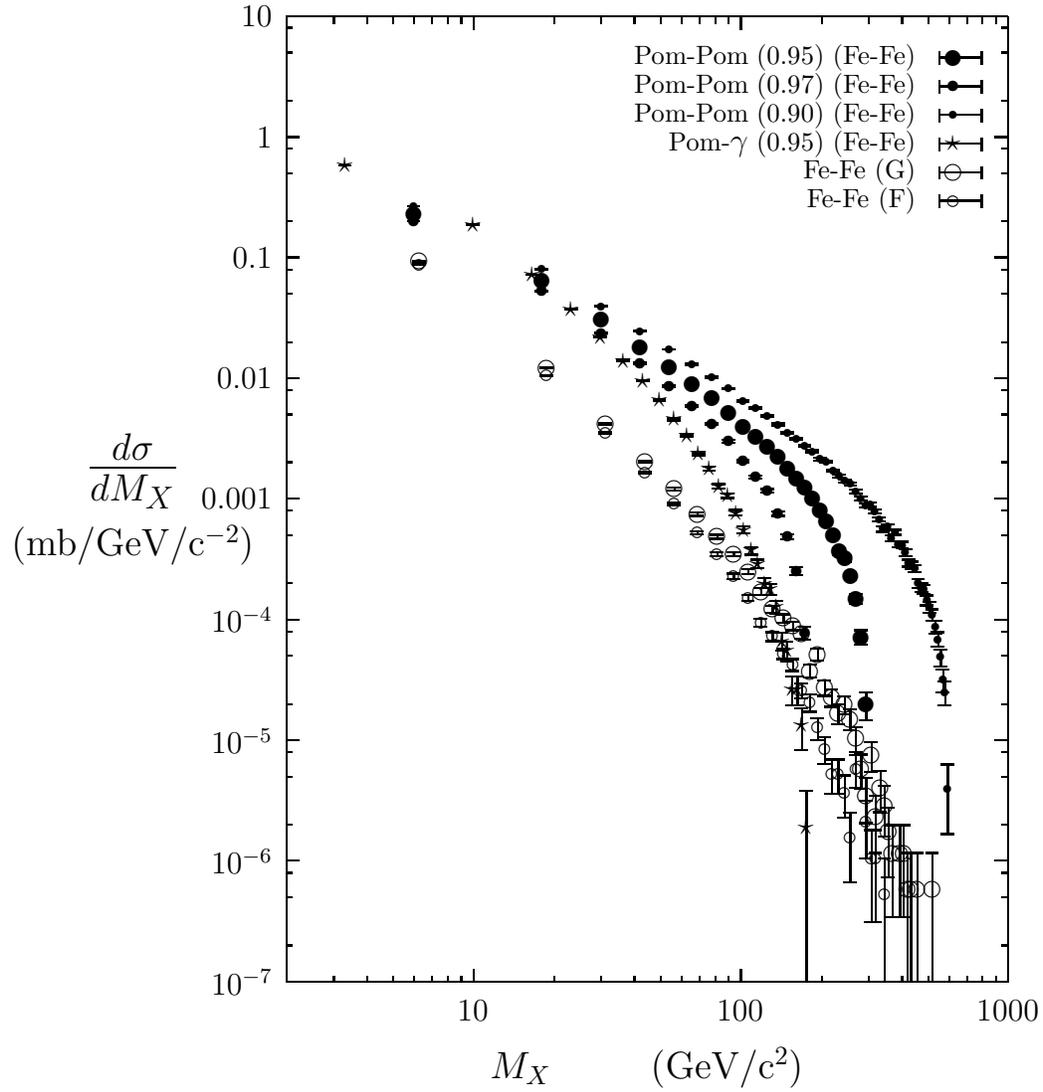}
 \vspace*{1cm}
\caption{
As Fig.~\protect\ref{hipssmblo} but for heavy ion reactions
 Fe-Fe.
\label{hifessmblo}
}
\end{figure}

\begin{figure}[thb] \centering
\hspace*{0.25cm}
\input{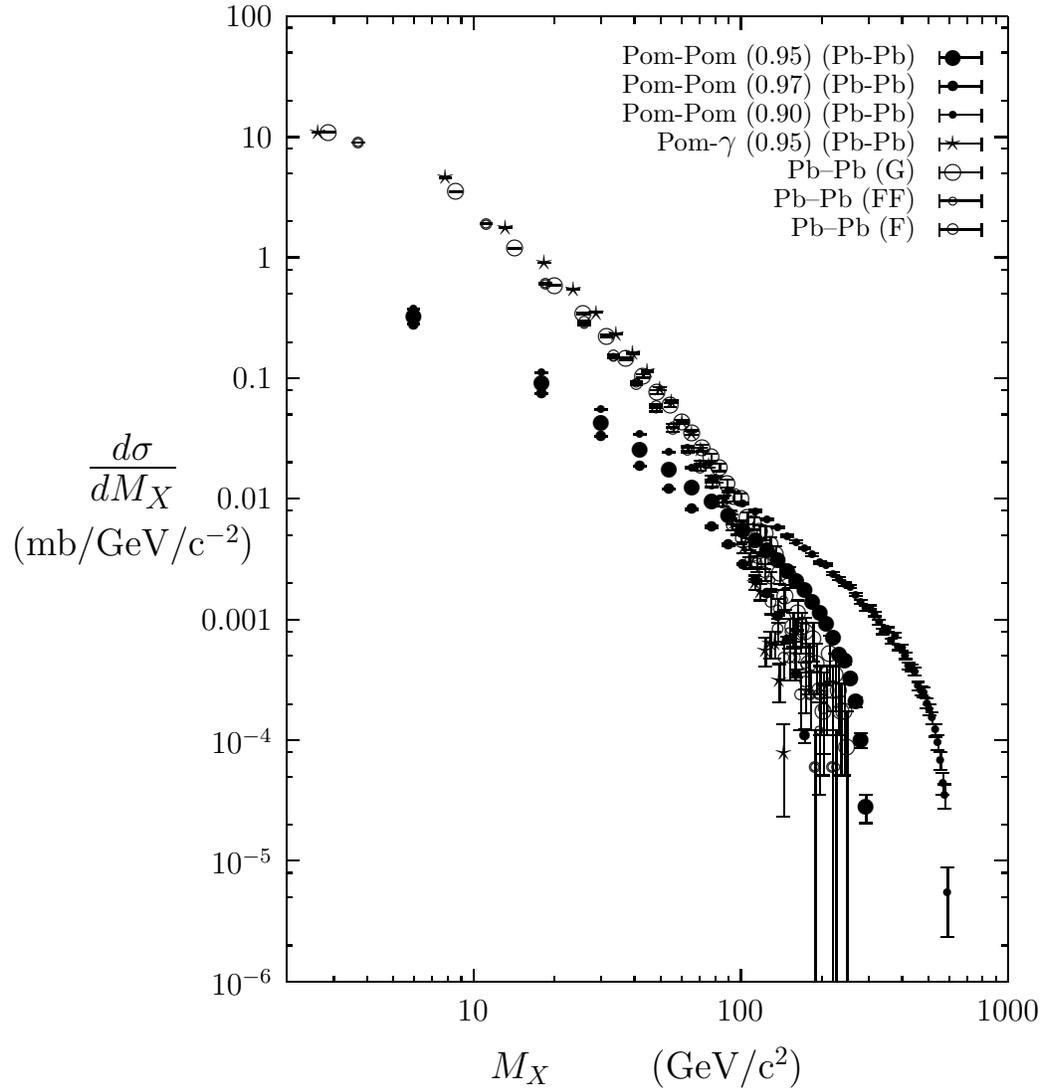}
 \vspace*{1cm}
\caption{
As Fig.~\protect\ref{hipssmblo} but for heavy ion reactions
 Pb-Pb.
\label{hipbssmblo}
}
\end{figure}

In Figs.~\ref{hipssmblo}--\ref{hipbssmblo} we
compare the cross section for the
production of a hadronic cluster of invariant mass $M_{X}$ in
photon-photon collisions 
 with the
corresponding cross section for the double-pomeron and
photon-pomeron reactions.
The calculation was done for $pp$ collisions and for 
heavy ion collisions O-O,  Ca-Ca, Fe-Fe, Ag-Ag, and Pb-Pb 
at the energy of the LHC
($\sqrt s $ = 6  A TeV).
For the photon-photon collisions we show the results using
the form factor approximation (F) and
the semi-classical, geometric approximation (G)
to calculate the heavy ion photon flux.
For the Pb-Pb reaction we use in fact two different 
form factors, (F) the geometric approximation to the form
factor (see Eq.~(\ref{geom-ff})) and
(FF), the Gaussian approximation. Both approximations lead to
nearly identical cross sections $d\sigma $/$dM_{X}$, therefore, we
present for all other reactions only the (F) cross section. 
The double-pomeron cross sections are given for three different
kinematical cuts ($M_{cd} >$ 2 GeV/c${}^2$, $c =$ 0.90, 0.95 and
0.97). The photon-pomeron cross sections are given for only one
setting of the kinematical cuts ($M_{\gamma\Pom} > $ 2
GeV/$c^2$ and c = 0.95). 
Of course, these cuts depend on the experimental setup of
such an experiment,
which is not known at present, but any cuts are easy to
apply to our Monte Carlo events.

Even given the considerable uncertainty in the cross sections
for CD and photon-pomeron collisions, 
the conclusions from Figs.~\ref{hipssmblo}--\ref{hipbssmblo} 
are rather
obvious. The study of the pure two-photon reaction without 
  background from central  diffraction and photon-pomeron
  collisions is  not
 possible. 
 In proton-proton collisions  and collisions of
 light and medium heavy ions the central  diffraction
 reaction dominates. For heavy ion collisions the
 photon-pomeron cross section is comparable to the
 photon-photon cross section. 
 Many interesting particles might be produced
 and studied in photon-photon photon-pomeron 
 as well as pomeron-pomeron
 collisions. If a reaction is to be studied using pomeron-pomeron
 collisions, then the best results should be obtained 
 in $pp$ collisions, where the highest luminosity can be obtained. 

\subsection{Rapidity distributions }

\begin{figure}[thb] \centering
\hspace*{0.25cm}
\input{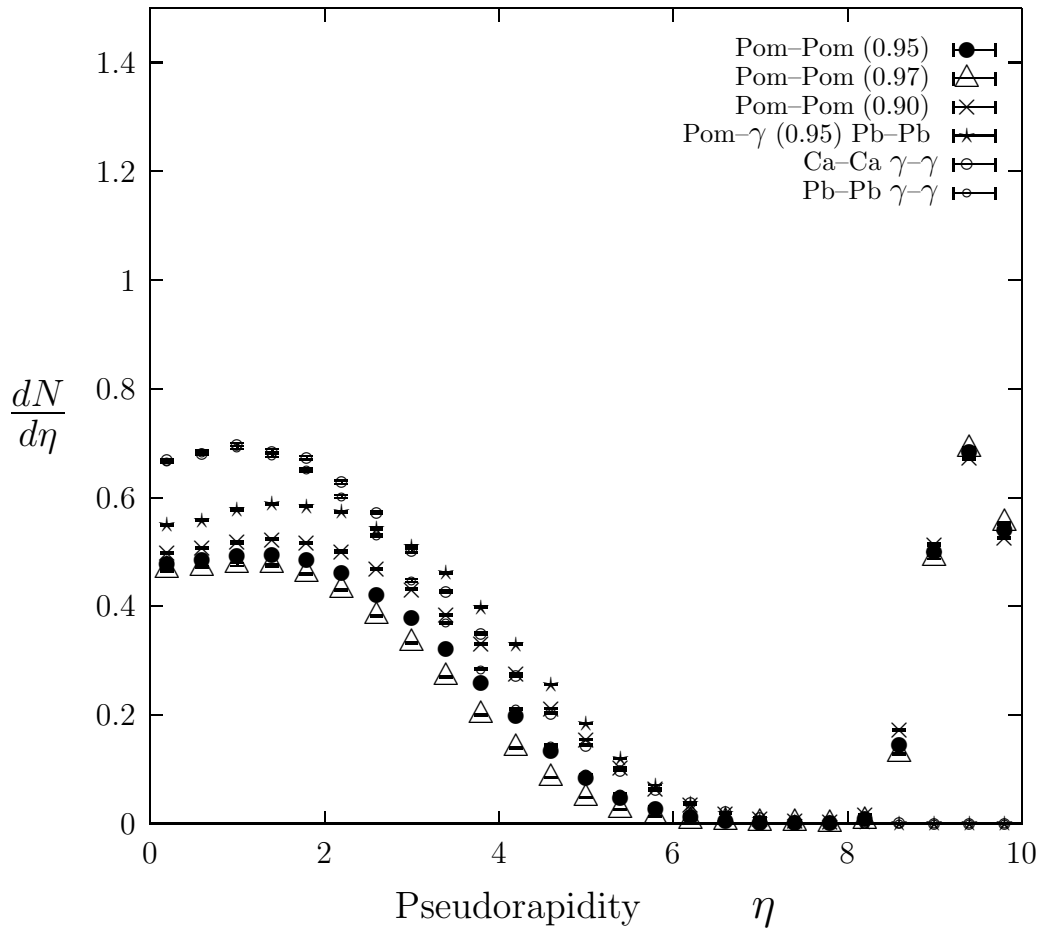}
 \vspace*{1cm}
\caption{
The pseudorapidity distribution of charged hadrons in photon-photon
collisions in Pb-Pb and Ca-Ca heavy ion collisions (only the
hadrons produced in the central cluster of particles are
included in the histogram) compared to the corresponding
distribution in central  diffraction in $pp$ collisions
(in this case also the scattered original protons are included
in the histogram) and to the corresponding distribution in
photon-pomeron collisions in Pb-Pb collisions (in this case
again only the hadrons in the central cluster are included in the
histogram). 
\label{higgpopo95etach}
}
\end{figure}

In order to demonstrate that the three reactions studied lead to very
similar signatures of the events we present in 
Fig.~\ref{higgpopo95etach} pseudorapidity distributions of the
produced hadrons.
The Figure shows
the distribution of hadrons in photon-photon
reactions  in Pb-Pb and Ca-Ca heavy ion collisions (only the
hadrons produced in the central cluster of particles are
included in the histogram).
The distribution in CD are presented for
$pp$ collisions and 
in this case we include also the scattered incoming protons 
into the histogram.  The distribution for photon-pomeron
collisions is presented for Pb-Pb collisions and includes only
the hadrons from the central cluster.

With the photon fluxes used and with the kinematical cuts for
the diffractive reactions, we obtain in all cases two
large rapidity gaps between the central cluster of hadrons and
the scattered original protons or heavy ions. The distributions
given in
Fig.~\ref{higgpopo95etach}  represent the average pseudorapidity
distributions averaged over many  collisions and the mass
spectrum given in Figs.~\ref{hipssmblo} to \ref{hipbssmblo}.
 In each single
event the pseudorapidity distribution corresponding to the
central cluster will be less wide than the average and it will
in general not be in the center of the nucleon-nucleon CMS.


\section{Conclusions and summary}

Photon-photon collisions as well as central  diffraction and
photon-pomeron collisions 
are very interesting reaction channels at proton-proton and 
heavy ion colliders in the TeV energy range.

Here we compare the cross sections of the three  channels for the
production of hadronic systems of given invariant mass $M_{X}$.
The three reaction channels have very similar experimental
signatures, a central cluster of produced particles and two
large rapidity gaps.

The cross section for CD  and photon-pomeron collisions 
is still subject to
large uncertainties (as large as a factor of three) 
in the TeV energy region.

We find for proton-proton and light up to  medium-heavy ion
reactions that the central  diffraction cross section 
dominates the two photon  and photon-pomeron cross sections. 
In collisions of the
heaviest ions like Pb-Pb the photon-pomeron channel and the 
photon-photon channel are of comparable magnitude,
however at very large masses CD can be larger
than the other two channels depending on the experimental
cuts.

The conclusion is obvious. Central double-pomeron processes are
best studied in $pp$ collisions, where the largest luminosities
are obtained, this compensates the rise of the double-pomeron
cross section in heavy ion reactions.
If photon-photon reactions are to be studied, 
then it should be the best 
to use the heaviest ions available at future colliders but even
then the background from photon-pomeron collisions will be roughly of
the same size as the signal.

\vspace*{1cm}
{\bf Acknowledgements}\\
Discussions with K.~Eggert and S.~Roesler are gratefully
acknowledged.
The authors thank K.~Hencken for providing them the
code \cite{Hencken95a} to calculate photon fluxes in
heavy ion collisions using the semi-classical approximation.
One of the authors (R.E.) was supported by
the Deutsche Forschungsgemeinschaft under contract No. Schi
422/1-2. One of the authors (J.R.) was supported by the
Direccion General de Politicia Cientifica of Spain.

\begin{appendix}

\section{Appendix:Photon flux calculation\label{photon-flux}}

It is convenient to define the luminosity function
for the photon flux in hadron-hadron
scattering by
\begin{equation}
\frac{d L}{dy_1 dP_1^2 dy_2 dP_2^2 } = f(y_1,p_1^2;y_2,p_2^2) 
\Theta(s_{1,2} - s_{\rm min}).
\label{luminosity-gam-gam-gen}
\end{equation}
Here, $p_1$ and $p_2$ are the four momenta of the photons
forming the subsystem. 
$p_1^2 = - P_1^2$ and $p_2^2 = - P_2^2$ are the
photon virtualities. 
The variables $y_1$ and $y_2$ denote approximately the energy
fractions taken by the photons from the initial hadrons as
explained below.
The Heavyside function in (\ref{luminosity-gam-gam-gen}) restricts the invariant
mass of the system formed by the four momenta $p_1$ and $p_2$ to allow
the application of the model. 
To obtain the cross sections for the processes mentioned above, the luminosity
function is folded with the $\gamma\gamma$ cross section.

\begin{figure}[thb] \centering
\hspace*{0.25cm}
  \psfig{file=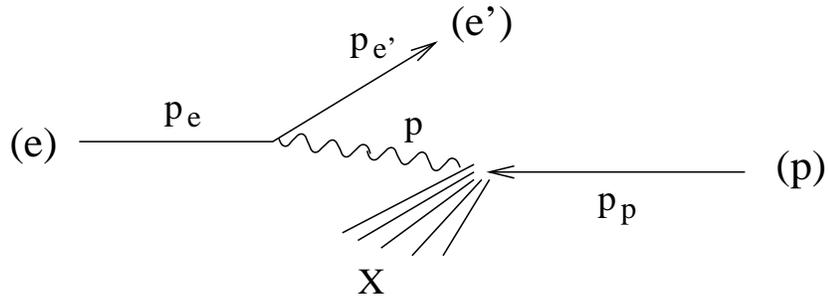,width=110mm}
 \vspace*{1cm}
\caption{
\label{ep-flx} 
Diagram of $ep$ scattering via one-photon exchange.}
\end{figure}

As a first step, photon emission off a pointlike particle is discussed
using as example the kinematics of $ep$ scattering,
shown in Fig.~\ref{ep-flx}. To characterize 
deep-inelastic scattering, we use the variables $x$ and $y$
\begin{equation}
x = \frac{P^2}{2 (p_p\cdot p)} \hspace*{2cm} 
y = \frac{(p\cdot p_p)}{(p_e\cdot p_p)}
\end{equation}
where $x$ denotes Bjorken's scaling variable.
Then, the differential cross section for $ep$ scattering via photon exchange
can be written in terms of the structure functions $F_1(x,P^2)$ and 
$F_2(x,P^2)$
\begin{eqnarray}
\frac{d\sigma^{ep}}{dy dP^2} &=& \frac{4 \pi \alpha^2_{\rm em}}{P^4}
\Bigg\{ x y \left(1-\frac{2 m_e^2}{P^2}\right)\;F_1(x,P^2)
\nonumber\\ & &\hspace*{-1.2cm}+ 
\frac{1}{y}\left(
1-y-\frac{m_p^2 P^2}{((p_e\cdot p_p)^2-m_e^2-m_p^2)^2} \right)\; F_2(x,P^2) 
\Bigg\}.
\nonumber\\
\label{gen-flux}
\end{eqnarray}
Here, $\alpha_{\rm em}$ denotes the fine structure constant and $m_e$ is
the electron mass.
In the limit of high collision energies, terms proportional to
$m_p^2/(p_e\cdot p_p)^2$ and $m_e^2/(p_e\cdot p_p)^2$ can be neglected.
Using the optical theorem, the structure functions can be related to
the total $\gamma p$ cross sections for virtual photons
with transverse polarization (helicity $\pm 1$) $\sigma^{\gamma p}_T$ 
and scalar polarization (helicity 0) $\sigma^{\gamma p}_S$ \cite{Budnev75}
\begin{eqnarray}
F_1(x,P^2) &=& \frac{(p_p\cdot p) (1-x)}{8 \pi^2 \alpha_{\rm em}} 
\sigma^{\gamma p}_T\\
F_2(x,P^2) &=& 2 \frac{P^2 (1-x)(\sigma^{\gamma p}_T + \sigma^{\gamma p}_S)}{
8 \pi^2 \alpha_{\rm em} 
(1-m_p^2 P^2/(p_p\cdot p)^2))}\ .
\end{eqnarray}
For small values of $P^2$, (\ref{gen-flux}) simplifies to
\begin{equation}
\frac{d\sigma^{ep}}{dy dP^2} = \frac{\alpha_{\mbox{\scriptsize em}}
}{2 \pi P^2} \left( \frac{1+(1-y)^2}{y} - 2 m_e^2 y \frac{1}{P^2}
\right) \sigma^{\gamma p}_T\ .
\end{equation}
It should be emphasized that in the high-energy limit the flux of
weakly virtual photons can be factorized out and is independent of 
the second scattering particle, which allows to 
introduce the generic photon flux function for bremsstrahlung
\begin{equation}
f_{\gamma,e}(y,P^2) = \frac{\alpha_{\mbox{\scriptsize em}}}{2 \pi P^2}
\left( \frac{1+(1-y)^2}{y} - 2 m_e^2 y \frac{1}{P^2}
\right)\ .
\label{bremsst}
\end{equation}

Due to the complex structure of the charge distribution $\rho(\vec x)$
in hadrons, 
several approximations and assumptions are necessary to calculate 
the flux of weakly virtual photons. The
approaches in literature can be subdivided into \cite{Baur88b}:
(i) methods using charge
form factors for the hadrons \cite{Grabiak89a,Drees89a}
and (ii) methods using the semi-classical methods and geometrical
interpretations on the basis of the impact parameter representation
(for example \cite{Papageorgiu90a,Baur88a,Baron93a,Hencken95a}).

In the form factor approach, Eq.~(\ref{bremsst})
can be used directly. The effects due to the finite charge
space-distribution can be included by substituting 
\begin{equation}
\alpha_{\rm em} \longrightarrow Z^2 \alpha_{\rm em} | F(p^2) |^2
\end{equation}
for each colliding hadron where $Z$ denotes the electric charge number.
The weak point on this approach are
the almost unknown elastic form factors $F(p^2)$ for heavy ions.
The simplest assumption for the heavy ion elastic form factor
is motivated by the geometrical interpretation:
In the classical picture one should only 
consider photons having an impact parameter $\vec B$
relative to the hadron greater than the transverse hadron 
size $R\approx 1.2 {\rm fm}\; A^{1/3}$.
With $P^2 \sim 1/\vec{B}^2$ follows
\begin{equation}
F(p^2) = \int d^3x\; \rho(\vec x)\; e^{i \vec p \cdot \vec x} = 
\left\{\begin{array}{r@{\quad  \quad}l}
1 \ ,  & - p^2 < 1/R^2 \\  & \\
0 \ , & - p^2 \ge 1/R^2
\end{array}\right. 
\label{geom-ff}
\end{equation}
More realistic parametrizations of the elastic form factor can be found in
literature \cite{Barret77a}.  For example, the form factor of ${}^{206}$Pb
can be parametrized by a Gaussian distribution
$F(p^2) = \exp( p^2/Q_0^2)$ with
$Q_0 \approx 55 - 60$ MeV as used in Ref.~\cite{Drees89a}.

\begin{figure}[thb] \centering
\hspace*{0.25cm}
  \psfig{file=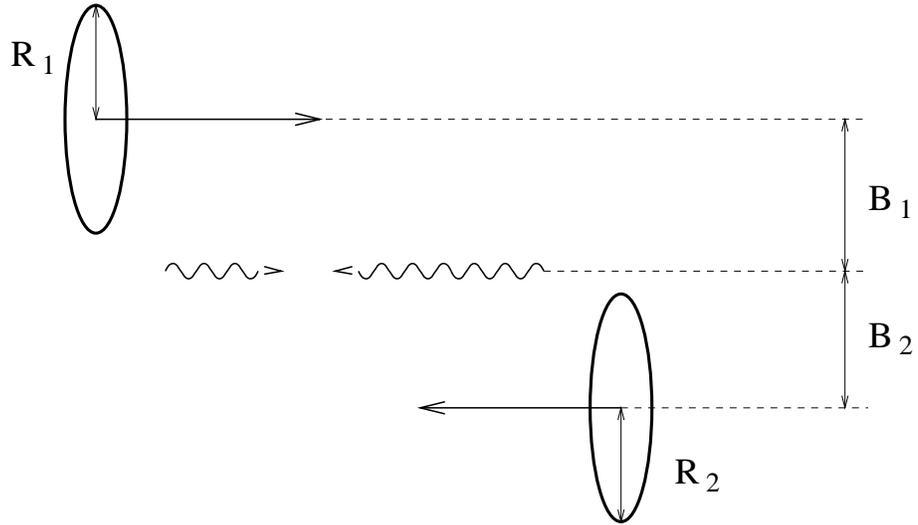,width=120mm}
 \vspace*{1cm}
\caption{
\label{hion-flx} Semi-classical model of the photon-photon scattering
in hadron-hadron interactions.}
\end{figure}

The basis of the semi-classical (geometrical) methods is the 
fact that a fast-moving charged particle develops a magnetic field almost
of the same size as the electric field. This can be described by photons 
moving parallel to the particle at an impact parameter $\vec B$ 
(see Fig~\ref{hion-flx}).

The number of equivalent photons is given by \cite{Jackson63}
\begin{eqnarray}
f(y,\vec B) &=&
\nonumber\\
& &\hspace*{-1cm}\frac{Z^2\alpha_{\rm em}}{\pi^2}
(m y)^2 \frac{1}{y} \left[ K_1^2(m |\vec B| y) +
\frac{m^2}{E^2} K_0^2(m |\vec B| y) \right],
\end{eqnarray}
where $K_0$ and $K_1$ denote the modified Bessel functions, $E$ and $m$ are
the energy and the mass of the hadron (heavy ion), respectively. 
Since the virtualities of the photons are neglected, the photon energy 
is given directly by $\omega = y E$.
For elastic heavy ion scattering, the impact parameter of the
equivalent photons is restricted to $|\vec B| > R$.
The total photon flux follows from
\begin{eqnarray}
f(y) &=& 2 \pi \int_R^{\infty} f(y,\vec B) B dB 
\\
&=& \frac{2}{y} 
\frac{Z^2\alpha_{\rm
em}}{\pi} \left[ \xi K_0(\xi)K_1(\xi) -
\frac{\xi^2}{2}\left(K_1^2(\xi)-K_0^2(\xi)\right)\right]
\nonumber\\
\label{class-flux-geom}
\end{eqnarray}
with $ \xi = m R y$.
The transverse distance between the particles should be
larger than the transverse sizes of the particles to make sure that the 
particles do not interact hadronically 
which leads to the luminosity function 
\cite{Papageorgiu90a,Baron93a,Hencken95a} 
\begin{eqnarray}
\frac{d L}{dy_1 dy_2} &=& \int_{|\vec B_1| > R_1} \int_{|\vec B_2| > R_2}
f(y_1,\vec B_1) f(y_2,\vec B_2)
\nonumber\\
& &\hspace*{-1.5cm}\times\Theta(s_{1,2} - s_{\rm min})\;
\Theta( | \vec{B}_1 - \vec{B}_2 |-(R_1 + R_2))
\;d^2 B_1 d^2 B_2\ .
\nonumber\\
\label{luminosity-gam-gam-geom}
\end{eqnarray}

\begin{figure}[thb] \centering
\hspace*{0.25cm}
  \psfig{file=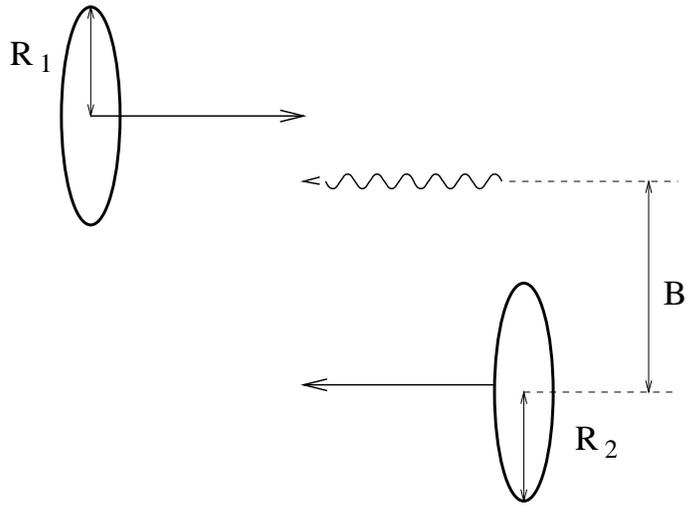,width=90mm}
 \vspace*{1cm}
\caption{
\label{pho-pom} Semi-classical model of the photon-hadron scattering
in hadron-hadron interactions. To obtain a model for photon-pomeron 
scattering, only single diffractive events where the photon 
dissociates are considered.}
\end{figure}

In the case of photon-pomeron scattering one has to consider photon-hadron 
(photon-heavy ion) scattering. The flux function reads (for example, for
the electromagnetic interaction of particle $B$)
\begin{equation}
\frac{d L}{dy_2} = \int_{|\vec B_2| > R} 
f(y_2,\vec B_2) \Theta(s_{1,2} - s_{\rm min})\ .
\end{equation}
Here, the radius $R$ is the minimum impact parameter of the photon that the 
heavy ions do not overlap in transverse space as shown in 
Fig.~\ref{pho-pom}. A value of 
$R=R_1+R_2$ would satisfy this condition, however,
since diffractive processes are mainly peripheral processes \cite{Faessler93a}, 
there may be 
also contributions with $R_2 < |\vec{B}| < R_1+R_2$. The results shown here
were obtained using $R=R_1+R_2$. Lowering the impact parameter
cutoff to $R=R_2$ increases the photon flux approximately by 20\%.

\end{appendix}
 
\vspace{1cm}
%


\begin{thebibliography}{10}

\bibitem{Baur88a}
C.~A. Bertulani and G.~Baur:
\newblock Phys.\ Rep.\ 163 (1988) 181

\bibitem{Grabiak89a}
M.~Grabiak, B.~M\"uller, W.~Greiner, G.~Soff  and P.~Koch:
\newblock J.\ Phys.\ G15 (1989) L25

\bibitem{Papageorgiu90a}
E.~Papageorgiu:
\newblock Phys.\ Lett.\ B250 (1990) 155

\bibitem{Vidovic93a}
M.~Vidovi\'c, M.~Greiner, C.~Best  and G.~Soff:
\newblock Phys.\ Rev.\ C47 (1993) 2308

\bibitem{Hencken95a}
K.~Hencken, D.~Trautmann  and G.~Baur:
\newblock Z.\ Phys.\ C68 (1995) 473

\bibitem{Drees89a}
M.~Drees, J.~Ellis  and D.~Zeppenfeld:
\newblock Phys.\ Lett.\ B223 (1989) 454

\bibitem{Papageorgiu89b}
E.~Papageorgiu:
\newblock Phys.\ Rev.\ D40 (1989) 92

\bibitem{Greiner91a}
M.~Greiner, M.~Vidovi\'c, J.~Rau  and G.~Soff:
\newblock J.\ Phys.\ G17 (1991) L45

\bibitem{Greiner93a}
M.~Greiner, M.~Vidovi\'c  and G.~Soff:
\newblock Phys.\ Rev.\ C47 (1993) 2288

\bibitem{Papageorgiu95a}
E.~Papageorgiu:
\newblock Phys.\ Lett.\ B352 (1995) 394

\bibitem{Ohnemus94}
J.~Ohnemus, T.~F. Walsh  and P.~M. Zerwas:
\newblock Phys.\ Lett.\ B328 (1994) 369

\bibitem{ALICETP}
{ALICE Collaboration}:
\newblock ALICE technical proposal,
\newblock CERN report CERN/LHCC/95--71,  1995

\bibitem{Ingelman85}
G.~Ingelman and P.~E. Schlein:
\newblock Phys.\ Lett.\ B152 (1985) 256

\bibitem{Bonino88}
UA8 Collab.:  R.~Bonino et~al.:
\newblock Phys.\ Lett.\ B211 (1988) 239

\bibitem{Brandt92}
UA8 Collab.:  A.~Brandt et~al.:
\newblock Phys.\ Lett.\ B297 (1992) 417

\bibitem{Schaefer90}
A.~Schaefer, O.~Nachtmann  and R.~Schoepf:
\newblock Phys.\ Lett.\ B249 (1990) 331

\bibitem{Bialas91}
A.~Bialas and P.~V. Landshoff:
\newblock Phys.\ Lett.\ B256 (1991) 540

\bibitem{Fletcher93}
R.~S. Fletcher and T.~Stelzer:
\newblock Phys.\ Rev.\ D48 (1993) 5163

\bibitem{Ellis90b}
J.~Ellis and P.~Salati:
\newblock CERN Report CERN TH--5693,
\newblock 1990

\bibitem{Streng86a}
K.~H. Streng:
\newblock Phys.\ Lett.\ 166B (1986) 443

\bibitem{Bialas92}
A.~Bialas and W.~Szeremeta:
\newblock Phys.\ Lett.\ B296 (1992) 191

\bibitem{Bialas94a}
A.~Bialas and R.~Janik:
\newblock Z.\ Phys.\ C62 (1994) 487

\bibitem{Heyssler96}
M.~Heyssler:
\newblock Diffractive heavy flavor production at the TEVATRON and the LHC,
\newblock DTP/96/10,
\newblock 1996

\bibitem{Natale94a}
A.~A. Natale:
\newblock Mod.\ Phys.\ Lett.\ A22 (1994) 2075

\bibitem{Natale95a}
A.~A. Natale:
\newblock Phys.\ Lett.\ B362 (1995) 177

\bibitem{Capella87b}
A.~Capella, J.~Tran Thanh~Van  and J.~Kwiecinski:
\newblock Phys.\ Rev.\ Lett.\ 58 (1987) 2015

\bibitem{Aurenche92a}
P.~Aurenche, F.~W. Bopp, A.~Capella, J.~Kwiecinski, M.~Maire, J.~Ranft  and
  J.~Tran Thanh~Van:
\newblock Phys.\ Rev.\ D45 (1992) 92

\bibitem{Bopp94a}
F.~W. Bopp, R.~Engel, D.~Pertermann  and J.~Ranft:
\newblock Phys.\ Rev.\ D49 (1994) 3236

\bibitem{Engel95a}
R.~Engel:
\newblock Z.\ Phys.\ C66 (1995) 203

\bibitem{Engel95d}
R.~Engel and J.~Ranft:
\newblock Hadronic photon-photon collisions at high energies,
\newblock preprint ENSLAPP-A-540/95,
\newblock (hep-ph/9509373),  1995

\bibitem{Engel95c}
R.~Engel, J.~Ranft  and S.~Roesler:
\newblock Phys.\ Rev.\ D52 (1995) 1459

\bibitem{Engelpopo}
R.~Engel and J.~Ranft:
\newblock {to be published},
\newblock 1996

\bibitem{Aurenche94a}
P.~Aurenche, F.~W. Bopp, R.~Engel, D.~Pertermann, J.~Ranft  and S.~Roesler:
\newblock Comp.\ Phys.\ Commun.\ 83 (1994) 107

\bibitem{Engel94c}
R.~Engel:
\newblock Photoproduction within the Dual Parton Model,
\newblock Talk given at XXIXth Rencontre de Moriond, in Proceedings of the
  XXIXth Rencontre de Moriond, Page 321, ed.\ by J.~Tr\^an Thanh V\^an, Edition
  Frontieres, Gif-sur-Yvette,  1994

\bibitem{Engel95b}
R.~Engel:
\newblock Multiparticle Photoproduction within the two-component Dual Parton
  Model,
\newblock in preparation,  1996

\bibitem{GRV92a}
M.~Gl\"uck, E.~Reya  and A.~Vogt:
\newblock Phys.\ Rev.\ D45 (1992) 3986

\bibitem{GRV92b}
M.~Gl\"uck, E.~Reya  and A.~Vogt:
\newblock Phys.\ Rev.\ D46 (1992) 1973

\bibitem{Abramovski73}
V.~A. Abramovski, V.~N. Gribov  and O.~V. Kancheli:
\newblock Yad.\ Fis.\ 18 (1973) 595

\bibitem{TerMartirosyan73}
K.~A. Ter-Martirosyan:
\newblock Phys.\ Lett.\ B44 (1973) 377

\bibitem{Hahn90}
K.~Hahn and J.~Ranft:
\newblock Phys.\ Rev.\ D41 (1990) 1463

\bibitem{Capella80b}
A.~Capella, U.~Sukhatme, C.~I. Tan  and J.~Tran Thanh~Van:
\newblock Z.\ Phys.\ C10 (1980) 249

\bibitem{Kaidalov82a}
A.~B. Kaidalov:
\newblock Phys.\ Lett.\ B116 (1982) 459

\bibitem{Capella96a}
A.~Capella, A.~Kaidalov, C.~Merino, D.~Pertermann  and J.~Tran Thanh~Van:
\newblock Phys.\ Rev.\ D53 (1996) 2309

\bibitem{Bengtsson87}
H.~U. Bengtsson and T.~Sj\"ostrand:
\newblock Comp.\ Phys.\ Commun.\ 46 (1987) 43

\bibitem{Good60}
M.~L. Good and W.~D. Walker:
\newblock Phys.\ Rep.\ 120 (1960) 1854

\bibitem{Kaidalov79}
A.~B. Kaidalov:
\newblock Phys.\ Rep.\ 50 (1979) 157

\bibitem{Roesler93}
S.~Roesler, R.~Engel  and J.~Ranft:
\newblock Z.\ Phys.\ C59 (1993) 481

\bibitem{Capella74}
A.~Capella and J.~Kaplan:
\newblock Phys.\ Lett.\ B52 (1974) 448

\bibitem{Capella76}
A.~Capella, J.~Kaplan  and J.~Tran Thanh~Van:
\newblock Nucl.\ Phys.\ B105 (1976) 333

\bibitem{Gotsman93}
E.~Gotsman, E.~M. Levin  and U.~Maor:
\newblock Phys.\ Rev.\ D49 (1994) 4321

\bibitem{Aid95b}
H1 Collab.:  S.~Aid et~al.:
\newblock Z.\ Phys.\ C69 (1995) 27

\bibitem{Chew74}
D.~M. Chew and G.~F. Chew:
\newblock Phys.\ Lett.\ 53B (1974) 191

\bibitem{Kaidalov74a}
A.~B. Kaidalov and K.~A. Ter-Martirosyan:
\newblock Nucl.\ Phys.\ B75 (1974) 471

\bibitem{Berera95a}
A.~Berera and J.~C. Collins:
\newblock Double pomeron jet cross sections,
\newblock report PSU/TH/162,
\newblock (hep-ph/9509258),  1995

\bibitem{Pumplin93a}
J.~Pumplin:
\newblock Phys.\ Rev.\ D47 (1993) 4820

\bibitem{Gotsman95a}
E.~Gotsman, E.~M. Levin  and U.~Maor:
\newblock Phys.\ Lett.\ B353 (1995) 526

\bibitem{Armesto96a}
N.~Armesto and M.~A. Braun:
\newblock The pomeron-pomeron interaction in the perturbative QCD,
\newblock preprint US-FT/27-96,
\newblock (hep-ph/9606307),  1996

\bibitem{Blankenbecler81}
A.~Blankenbecler, A.~Capella, J.~Tranh Than~Van, C.~Pajares  and V.~A. Ramallo:
\newblock Phys.\ Lett.\ B 107 (1981) 106

\bibitem{Pajares81b}
C.~Pajares and V.~A. Ramallo:
\newblock Phys.\ Lett.\ B107 (1981) 238

\bibitem{Pajares85a}
C.~Pajares and V.~A. Ramallo:
\newblock Phys.\ Rev.\ D31 (1985) 2800

\bibitem{Braun91}
M.~A. Braun and C.~Pajares:
\newblock Nucl.\ Phys.\ A532 (1991) 678

\bibitem{Segre77}
E.~Segr\'e:
\newblock {\em Nuclei and particles}
\newblock Reading Mass Benjamin 1977

\bibitem{Ranft94b}
J.~Ranft and S.~Roesler:
\newblock Z.\ Phys.\ C62 (1994) 329

\bibitem{Faessler93a}
M.~A. Faessler:
\newblock Z.\ Phys.\ C58 (1993) 567

\bibitem{Brandt94a}
D.~Brandt, K.~Eggert  and A.~Morsch:
\newblock CERN Report CERN AT/94--05, LHC Note 264,
\newblock 1994

\bibitem{Budnev75}
V.~M. Budnev, I.~F. Ginzburg, G.~V. Meledin  and V.~G. Serbo:
\newblock Phys.\ Rep.\ 15C (1975) 181

\bibitem{Baur88b}
G.~Baur and C.~A. Bertulani:
\newblock Z.\ Phys.\ A330 (1988) 77

\bibitem{Baron93a}
N.~Baron and G.~Baur:
\newblock Phys.\ Rev.\ C48 (1993) 1999

\bibitem{Barret77a}
R.~C. Barret and D.~F. Jackson:
\newblock {\em Nuclear size and structure}
\newblock Clarendon Oxford 1977

\bibitem{Jackson63}
J.~D. Jackson:
\newblock {\em Classical electrodynamics}
\newblock John Wiley \& Sons New York 1963

\end{thebibliography}

\clearpage
\begin{table}[htb]
\caption{\label{table1}
Results for the effective $A^2_{\rm eff}$ in $AA$ collisions (second column)
and the effective
$A$-dependence in $hA$ collisions (third column) using 
$\sigma_{in}$ = 73 mb and Woods-Saxon nuclear densities.}
\vskip0.5cm
\renewcommand{\arraystretch}{1.5}
\begin{tabular}{|c|c|c|c|} \hline
Nucleus &  $A$  &  $A^2_{\rm eff}$ & $A_{\rm eff}$ \\ \hline
 O       & 16  & 4.94 & 2.36 \\
 Ca      & 40  & 6.21 & 2.92 \\
 Fe      & 56  & 6.76 & 3.16 \\
 Ag      & 108  & 8.00 & 3.71 \\
 Pb      & 208  & 9.52 & 4.39 \\
\hline
\end{tabular}
\end{table}
%
%

 \clearpage

 \clearpage

 \clearpage

 \clearpage

\begin{figure}[thb] \centering
\hspace*{0.25cm}
\input{hicassmblo.pic}
  \vspace*{1cm}
\caption{
as Fig.~\protect\ref{hipssmblo} but for heavy ion reactions
 Ca-Ca.
\label{hicassmblo}
}
\end{figure}
 \clearpage

 \clearpage

\begin{figure}[thb] \centering
\hspace*{0.25cm}
\input{hiagssmblo.pic}
 \vspace*{1cm}
\caption{
As Fig.~\protect\ref{hipssmblo} but for heavy ion reactions
 Ag-Ag.
\label{hiagssmblo}
}
\end{figure}
 \clearpage

 \clearpage

 \clearpage

 \clearpage

 \clearpage

\end{document}